\documentstyle[aps,pra]{revtex}

\begin{document}
\title{Pygmy resonances in artificial nuclei: far-infrared absorption by
electron-hole droplets}
\author{Roberto Capote\cite{capote}$^{1,2}$, Alain Delgado\cite{alain}$^2$and
Augusto Gonzalez\cite{augusto}$^{3,4}$}

\address{$^1$Dpto FAMN, Facultad de F\'{i}sica, Universidad de Sevilla,\\
Apdo 1065, Sevilla 41080, Espa\~{n}a \\
$^2$Centro de Estudios Aplicados al Desarrollo Nuclear,\\
Calle 30 No 502, Miramar, Habana 11300, Cuba \\
$^3$Instituto de Cibern\'{e}tica, Matem\'{a}tica y F\'{i}sica, \\
Calle E 309, Vedado, Habana 10400, Cuba \\
$^4$Departamento de F\'{i}sica, Universidad de Antioquia, AA 1040, \\
Medell\'{i}n, Colombia}
\date{Received: \today}
\maketitle

\begin{abstract}
The structure of E1 resonances is examined in a microscopic random phase
approximation calculation for neutral, symmetric, closed shell,
electron-hole systems in a quantum dot. The number of electron-hole pairs, $%
N $, is varied from $6$ to $42$. The ocurrence of small, but distinct, E1
peaks in the far infrared spectra located in the low energy tail of the
giant dipole resonance and consisting of highly coherent electron-hole
excitations is predicted. These pigmy resonances account for about 2 \% of
the dipole energy-weighted photoabsorption sum rule. A very weak dependence
of the average pigmy resonance energy on the number of electron-hole pairs
is found.

\end{abstract}

\medskip

A phenomena well studied in Nuclear Physics is the presence of giant
multipolar resonance modes related to collective excitations in nuclei.
Particularly, collective states corresponding to giant electric dipole
oscillations are known in nuclei long ago \cite{gdr,gdrrev}. They show
themselves as very high peaks in photoabsorption at energies $\sim
~80A^{-1/3}$ MeV, where A is the mass number. Both the excitation energy and
the total E1 strength are rather well understood from a collective\cite
{hydro} as well as from a microscopic point of view\cite{TD}. It is shown
that the giant dipole resonance (GDR) is a state coming from the splitting
of a degenerate set of $1^{-}$ states, which takes almost the whole strength
of the $1^{-}$ transition. Collective states like GDR have been extensively
studied, both experimentally and theoretically, in Atomic Physics. They are
at the base of giant and broad atomic resonances in the photoionisation
continuum \cite{Conn}. The photoabsorption cross section of metallic
clusters \cite{Conn,PWK91} exhibits also the dominance of a collective
oscillation mode at 2 - 3 eV \cite{BB94}. Theoretical calculations based on
the Random Phase Approximation (RPA) \cite{Conn,Lipp} are shown to reproduce
the resonance position and dipole strength distribution. In a previous paper%
\cite{PHYSEour}, we showed that the far infrared absorption probability in
neutral electron-hole systems confined in a quantum dot is also dominated by
a GDR. This result is mainly related to the existence of positively and
negatively charged particles, and not to a particular dot geometry. The
electron-hole system in a quantum dot shares many similarities with a real
nucleus: a self-consistent confinement potential, attractive and repulsive
channels in the residual interactions, etc. Therefore we want to verify if,
for such symmetric systems, we can also observe concentration of the E1
strength in a low excitation energy region, very similar to the so called
{\it pygmy resonances} in nuclear physics.

Recently, the study of electric dipole transitions from high-lying bound
states in nuclei gained much interest. The E1 strength distribution in
spherical nuclei near shell closures seems to display quite some fine
structure in this energy region and modulations of the GDR tail occur.
Experimental efforts have been made on $^{138}$Ba\cite{EXPBA}, $^{116,124}$Sn%
\cite{EXPSN}, $^{140}$Ce\cite{EXPCE} and the odd $^{89}$Y\cite{EXPY}, using
the photon scattering technique. A concentration of E1 strength turned up
clearly around 6.5 MeV in these nuclei. This pygmy resonance,
named as such in correspondence to the GDR, was first observed in heavier
nuclei in the Pb region\cite{EXPPB} at 5.5 MeV. Recent experiments on Ba and
Ce isotopes\cite{EXPBACE} , on $^{56}$Fe and $^{58}$Ni nuclei\cite{PRCC62},
and in the $N=50$ nuclei\cite{EXPN50} also show signatures of similar
phenomena. Finally, for the $^{92}$Mo nucleus a recent LINAC experiment \cite
{EXPMO} established the presence and the E1 character of this pygmy
resonance centered at 6.5 MeV. The observation of a fairly large dipole
strength$(B(E1))$ value, at relatively low excitation energy, might be due
to admixtures of the GDR into these low-lying $1^{-}$ states. \cite
{9OROS,10OROS} Theoretically, attention has been given to the observed local
concentration of E1 strength by, e.g., Iachello\cite{12OROS}, and Van
Isacker {\it et al}\cite{11OROS}. Iachello, discussing the examples of
clustering in nuclei and of permanent octupole deformation effects, has
suggested that with isospin as a local symmetry rather important
concentrations of E1 strength might well show up at low excitation energies.
Van Isacker {\it et al} suggest that nuclei with a reasonable neutron skin
may exhibit pygmy-E1 resonances below the GDR. The same phenomena was
explained by Oros {\it et al}\cite{OROS} using a two-group schematic random
phase approximation (RPA) model, suggesting that concentrations of strength
remain distributed among the unperturbed 1{\it p}-1{\it h} states when the
rest of the strength is pulled up into the GDR. A similar interpretation is
given by the Dubna group in a microscopic quasiparticle phonon model (QPM)%
\cite{13QPM}.

In this paper we will study the existence and behavior of pygmy resonances
in neutral quantum dot systems using a large harmonic oscillator basis ($%
\sim 20\hbar \omega $). A simplified model of disk-shaped dot is employed,
in which the lateral confinement is ideally parabolic \cite{WHFJ96}. $N$
electron-hole pairs are supposed to be created in the dot by means of, e.g.,
laser pumping with a typical $\sim $ 1 eV energy, as corresponding to a
semiconductor band gap. The created particles live hundreds of picoseconds
and even more \cite{BG93}. We will study the linear absorption of a second
far infrared, $\sim $ 10 meV, light wave by the $N$-pair cluster. The period
of this signal is $\sim 10^{-13}$s, thus the cluster may be considered as
stable when studying absorption. We will use the dipole approximation for
the interaction hamiltonian, and will consider in-plane light polarization..

In our model, there are only one electron and one hole bands, and $m_e=m_h$%
\cite{masses}. Electron-hole exchange will be neglected. The
second-quantized Hamiltonian for the two-dimensional motion of electrons and
holes is given elsewhere\cite{PHYSEour}. The important quantity is the ratio
of the Coulomb characteristic energy to the confinement energy of the dot,
denoted as $\beta =\sqrt{(\frac{me^4}{\kappa ^2\hbar ^2})/(\hbar \omega )}$,
where $\kappa $ is the dielectric constant, and $\omega $ is the dot
frequency. We consider the dot in the strong-confinement regime, where the
one-particle energy dominates the total energy. This is a good starting
picture for small dots with large $N$ values. We notice that both the
quasiplanar shape and the strong-confinement regime are common conditions
met in experiments on quantum dot luminescence\cite{experiments1}. We study
closed-shell quantum dots, thus we have a close analogy with near magic
nuclear systems where pygmy resonances were observed. The number of
electrons is equal to the number of holes, and it is given by the expression
$N=N_s(N_s+1)$, where $N_s$ is the last two-dimensional harmonic-oscillator
filled shell. Hartree-Fock shells are distorted harmonic oscillator shells.
The total angular momentum, total electron spin and total hole spin are all
equal to zero for the ground state.

The lowest excitations against the ground state are one-particle excitations
from the last filled shell to the next empty shell. The states with total
angular momentum projection, $M=\pm 1$, and electron and hole total spins, $%
S_e=S_h=0$, are the candidates among which we shall look for the GDR as well
as for pygmy resonances. To be definite, we will consider $M=1$. For $\beta
=0$ GDR and pygmy resonances are all degenerated in energy. Notice that
there are $N_s$ orbitals in the last shell, that is $2N_s$ electrons and,
thus, $2N_s$ wave functions with $M=1$. There are also $2N_s$ functions
corresponding to hole excitations. It makes a total of $4N_s$ degenerate
states. The Coulomb interaction breaks the degeneracy. The collective nature
of the resonances is manifested in the fact that the electric dipole
transition probability from the ground state is strongly enhanced \cite{TD}.
We will show that this property also holds for pygmy resonances using RPA
calculations.

In the RPA \cite{RR80}, we allow a general correlated ground state, $%
|RPA\rangle $, and the excited states are looked for in the form $\Psi
=Q^{\dagger }|RPA\rangle ,$ where the $Q^{\dagger }$ operator contains
``up'' and ``down'' transitions

\begin{equation}
Q^{\dagger}=\sum_{\sigma,\lambda}(X^e_{\sigma\lambda} e^{\dagger}_{\sigma}
e_{\lambda}+X^h_{\sigma\lambda} h^{\dagger}_{\sigma}
h_{\lambda}-Y^e_{\sigma\lambda} e^{\dagger}_{\lambda}
e_{\sigma}-Y^h_{\sigma\lambda} h^{\dagger}_{\lambda} h_{\sigma}).
\end{equation}

The $Y_{\sigma \lambda }$ coefficients are nonzero only for $\lambda $ and $%
\sigma $ satisfying $m_\lambda -m_\sigma =1$, where the $m_\lambda $ are
magnetic quantum numbers. The physical solutions annihilate the RPA ground
state $Q|RPA\rangle =0,$ and satisfy the normalization condition $%
1=\sum_{\sigma ,\lambda }\{(X_{\sigma \lambda })^2-(Y_{\sigma \lambda
})^2\}. $ The far infrared radiation to be absorbed is assumed to be
linearly polarized, with the electric field vibrating along the $y$
direction. The dipole matrix elements are computed from

\begin{eqnarray}
\langle \Psi |D| RPA\rangle &=& \sum_{\sigma,\lambda} \{
(X^h_{\sigma\lambda}-X^e_{\sigma\lambda}) \langle\sigma|y| \lambda\rangle
\nonumber \\
&+& (Y^h_{\sigma\lambda}-Y^e_{\sigma\lambda})\langle\lambda|y|\sigma\rangle
\}.
\end{eqnarray}

The Hartree-Fock energies and one-particle wave functions are obtained
selfconsistently by means of an iterative scheme\cite{PHYSEour}. Excitation
energies and dipole strength distributions were computed for $N$ ranging
from $6$ to $42$, and $\beta $ equal to $1$. Therefore, our Fermi levels
were always much lower than cut-off energy of the oscillator basis $\sim
20\hbar \omega$. The calculated dipole elements fulfill the energy-weighted
sum rule \cite{RR80}

\begin{eqnarray}
\sum_{\Psi}\Delta E~|\langle\Psi |D|RPA\rangle|^2= \sum_{\Psi_0}\Delta E_0
|\langle \Psi_0 |D|0 \rangle|^2=N/2.  \label{S1}
\end{eqnarray}

\noindent
where 0 indexes refer to the noninteracting $\beta =0$ case.

It should be stressed that for larger values of $\beta $, pairing becomes
very important \cite{GQRCR98}, and should be taken into account more
properly. We draw in Fig. \ref{fig1} the GDR and pygmy resonance excitation
energies at $\beta =1$ as a function of $N^{1/4}$. Notice the approximate
dependence $\Delta E_{GDR}-1\approx 0.7+0.4~N^{1/4}$ for a GDR energy and $%
\Delta E_{pygmy}-1\approx 0.7-0.1\ N^{1/4}$ for an average pygmy resonance
energy. At $\beta \ne 1$, the absolute energy shift $(\Delta E-1)$ has a
linear dependence on $\beta $. A remarkable feature is that the total number
of pygmy resonances is equal to $N_s-1$. Indeed, we checked that no pygmy
resonances are obtained for the case of two confined pairs $(N=2)$, where
the whole dipole strength goes to the GDR state.

It can be seen that the average pygmy resonance energy has a very weak
dependence on the number of electron-hole pairs $N$, confined in the system.
It is interesting to note that such weak dependence for average pygmy
resonance energy behavior is obtained in highly symmetric quantum dot
systems, equivalent to atomic nuclei with total isospin equal 0 $(N=Z)$.
This conclusion is in agreement with the results for $^{204,206,208}$Pb
nuclei\cite{PHYSLET2000}, where the comparison of available experimental
data below 6.5 MeV suggests that the E1 strength in this region can not be
attributed to an oscillation of the excess neutrons with respect to the
remaining nucleus core.

The similarity with the nuclear system is confirmed by inspecting the
experimental E1 strength distribution, measured up to endpoint energy of
6.75 MeV, in $^{204,206,208}$Pb nuclei\cite{PHYSLET2000}. The GDR energy for
these nuclear systems is about 13.5 MeV and the average energy for observed
pygmy resonances is about 6 MeV ($\sim 0.45$ times the GDR energy). The
dipole strength and the number of pygmy resonances increases when the
number of neutron pairs increase by 2 from $^{204}$Pb to $^{208}$Pb nucleus.
The results for closed shell quantum dots with $N=6,12,20,30$, and 42
confined electron-hole pairs at $\beta =1$ are presented in Fig. \ref{fig2}.
In analogy with the lead nuclei, the number of pygmy resonance states
increases when the total number of pairs is increased in the system. Thus,
the overall behavior of both systems is quite similar.

For the quantum dot systems, the fraction of the sum rule (\ref{S1})
accounted for the GDR state is close to 98 \%. The remaining 2 \% of the E1
strength lies at much lower energy (about 0.5-0.6 times the GDR energy)
corresponding to the pygmy resonance states, in analogy with the
behavior of the nuclear systems. Notice that only collective (highly
coherent) states, for which $X^e=-X^h$, are represented in our calculations
(see Fig. \ref{fig2}). The rest of the states show a threefold degeneracy
and give practically zero dipole matrix elements. Let us note that the
energy gap is increased ($\Delta E>1$) after the Coulomb interaction is
switched on. Energy of the collective dipole states increases when coulomb
interaction is considered. This is a signal that the overall interaction
among particles is repulsive.

We have shown that the far infrared absorption spectrum of neutral confined
systems of electrons and holes is dominated by collective dipole resonances,
including GDR as well as pygmy resonances. The latter account for
about 2\% of the energy-weighted photoabsorption sum rule, lying at average
energy about $0.5-0.6$ times the GDR energy. Calculations were done in a
two-band model with $m_e=m_h$ and a disk-shaped parabolic dot. The
qualitative conclusions are, however, expected to be valid for realistic
systems since they are mainly related to the existence of positively and
negatively charged particles in the system, which causes the enhancement of
dipole oscillations.

The authors acknowledge support from the Comite de Investigaciones de la
Universidad de Antioquia and from the Caribbean Network for Theoretical
Physics. We are grateful to Prof.M.Lozano for revision of the manuscript
and useful comments. One of the authors(R.C.) acknowledges support from the
Ministerio de Educaci\'{o}n y Cultura de Espa\~{n}a.

\newpage

\subsection{Figure captions:}

{\bf Figure 1.} GDR and {\it pigmy resonance} excitation energies at $\beta
=1$ as a function of the number of pairs, $N$, in the dot. Dashed and solid
line correspond to the linear fit for GDR and average {\it pigmy resonance}
energy as a function of $N$. The linear fit y=b*X+a parameters for GDR
energy are a=1,75+/-0,01 and b=-0,105+/-0,007 and for average {\it pigmy
resonance} energy are a=1,717+/-0,005 and b=0,394+/-0,003 correspondingly.

{\bf Figure 2.} Dipole matrix elements squared at $\beta =1$ as a function
of the number of electron-hole pairs, $N$.


\begin{references}
\bibitem[a]{capote}  Electronic address: rcapotenoy@yahoo.com

\bibitem[b]{alain}  Electronic address: gran@ceaden.edu.cu

\bibitem[c]{augusto}  Electronic address: agonzale@cidet.icmf.inf.cu

\bibitem{gdr}  A. de Shalit and H. Feschbach, {\it Theoretical Nuclear
Physics}, (John Wiley \& Sons, New York, 1974), Vol. I.

\bibitem{gdrrev}  Review articles and recent results can be found in {\it %
Electric and magnetic giant resonances in nuclei}, edited by J. Speth,
(World Scientific, Singapore, 1991).

\bibitem{hydro}  M. Goldhaber and E. Teller, Phys. Rev. {\bf 74}, 1046
(1948); H. Steinwedel and J. H. D. Jensen, Zeit. fur Naturforsch {\bf 5},
413 (1950).

\bibitem{TD}  J. P. Elliot and B. H. Flowers, Proc. Roy. Soc. (London) A
{\bf 242}, 57 (1957); A. M. Lane, {\it Nuclear Theory}, (Benjamin, New York,
1964); G. E. Brown, {\it Unified Theory of Nuclear Models}, (North Holland,
1964).

\bibitem{Conn}  Review articles can be found in {\it Correlations in
clusters and related systems}, edited by J. P. Connerade, (World Scientific,
Singapore, 1996).

\bibitem{PWK91}  S. Pollach, C. R. C. Wang and M. M. Kappes, J. Chem. Phys.
{\bf 94}, 2496 (1991).

\bibitem{BB94}  G. F. Bertsch and R. A. Broglia, {\it Oscillations in finite
quantum systems}, (Cambridge University Press, Cambridge, 1994).

\bibitem{Lipp}  E. Lipparini in {\it Many body theory of correlated electron
systems}, edited by M. I. Gallardo and M. Lozano, (World Scientific,
Singapore, 1998).

\bibitem{PHYSEour}  A. Delgado, L. Lavin, R. Capote and A. Gonzalez, Phys. 
E {\bf 8}, 345 (2000).

\bibitem{EXPBA}  R. D. Herzberg {\it et al}, Phys. Rev. C{\bf \ 60},
051307(R) (1999).

\bibitem{EXPSN}  K. Govaert {\it et al}, Phys. Rev. C{\bf \ 57}, 2229 (1998).

\bibitem{EXPCE}  R. D. Herzberg {\it et al}, Phys. Lett. B{\bf \ 390}, 49
(1997).

\bibitem{EXPY}  J. Reif et al, Nucl. Phys. A{\bf \ 620}, 1 (1997).

\bibitem{EXPPB}  R. M. Laszewski and P. Axel, Phys. Rev. C{\bf \ 19}, 342
(1979)

\bibitem{EXPBACE}  R. M. Laszewski, Phys. Rev. C{\bf \ 34}, 1114 (1986).

\bibitem{PRCC62}  F. Bauwens {\it et al}, Phys. Rev. C{\bf \ 62}, 024302
(2000).

\bibitem{EXPN50}  L. Cannel, Ph. D. Thesis, University of Illinois, 1986; K.
Wienhard {\it et al}, Z. Phys. A{\bf \ 302}, 185 (1981).

\bibitem{EXPMO}  F. Bauwens {\it et al}, in {\it Abstracts of the General
Scientific Meeting of the Belgian Physical Society, VUB Brussels, 1999}, p.
NP8.

\bibitem{9OROS}  A. Zilges, P. von Brentano and A. Richter, Z. Phys. A{\bf \
341}, 489 (1992).

\bibitem{10OROS}  K. Heyde and C. De Coster, Phys. Lett. B{\bf \ 393}, 7
(1997).

\bibitem{12OROS}  F. Iachello, Phys.Lett. B{\bf \ 160}, 1 (1985) and
references therein.

\bibitem{11OROS}  P. Van Isacker, M. A. Nagarajan and D. D. Warner, Phys.
Rev. C{\bf \ 45}, 13(R) (1992).

\bibitem{OROS}  A. M. Oros, K. Heyde, C. De Coster and B. Decroix, Phys.
Rev. C{\bf \ 57}, 990 (1998).

\bibitem{13QPM}  D. T. Khoa, V. Yu. Ponomarev and V. V. Voronov, Izv. Acad.
Nauk SSSR, Ser. Fiz. {\bf 48}, 1846 (1984); V. G. Soloviev, {\it Theory of
Atomic Nuclei: Quasiparticles and phonons}, (Institute of Physics, Bristol,
1992); V. G. Soloviev, Ch. Stoyanov and V. V. Voronov, Nucl. Phys. A 
{\bf 304}, 503 (1978).

\bibitem{WHFJ96}  A. Wojs, P. Hawrylak, S. Fafard, and L. Jacak, Phys. Rev. B
{\bf 54}, 5604 (1996).

\bibitem{BG93}  See, for example, the experimental reports in {\it Optics of
excitons in confined systems}, edited by G. Bastard and B. Gil, Journal de
Physique IV, Vol. 3, Colloque C3 (1993).

\bibitem{masses}  This unrealistic condition leads to simplifications and
makes easier the comparison with nuclei. In the commonly used GaAs, for
example, the ratio of in-plane masses is $m_e/m_{hh}\approx 0.6$. The mass
asymmetry will split the resonances.

\bibitem{experiments1}  E. Dekel, D. Gershoni, E. Ehrenfreund, J. M. Garcia
and P. M. Petroff, Phys. Rev. Lett. {\bf 80}, 4991 (1998).

\bibitem{RR80}  P. Ring and P. Schuck, {\it The nuclear many-body problem},
(Springer-Verlag, New-York, 1980).

\bibitem{GQRCR98}  B. Rodr\'{i}guez, A. Gonz\'{a}lez, L. Quiroga, R. Capote
and F. J. Rodr\'{i}guez, Int. Journ. Mod. Phys. B {\bf 14}, 71 (2000).

\bibitem{PHYSLET2000}  J. Enders {\it et al}, Phys. Lett. B 16220 (2000).
\end{references}
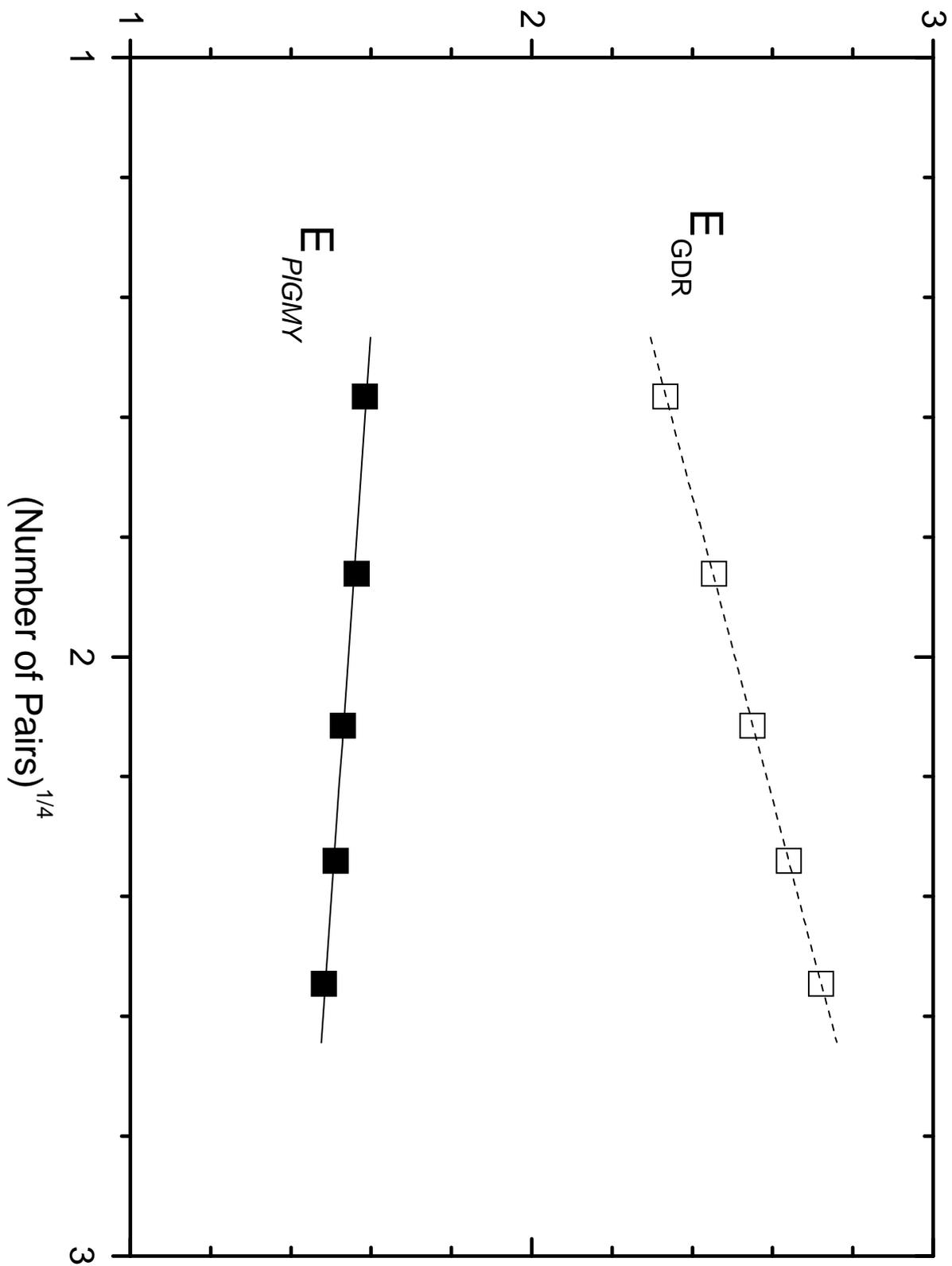
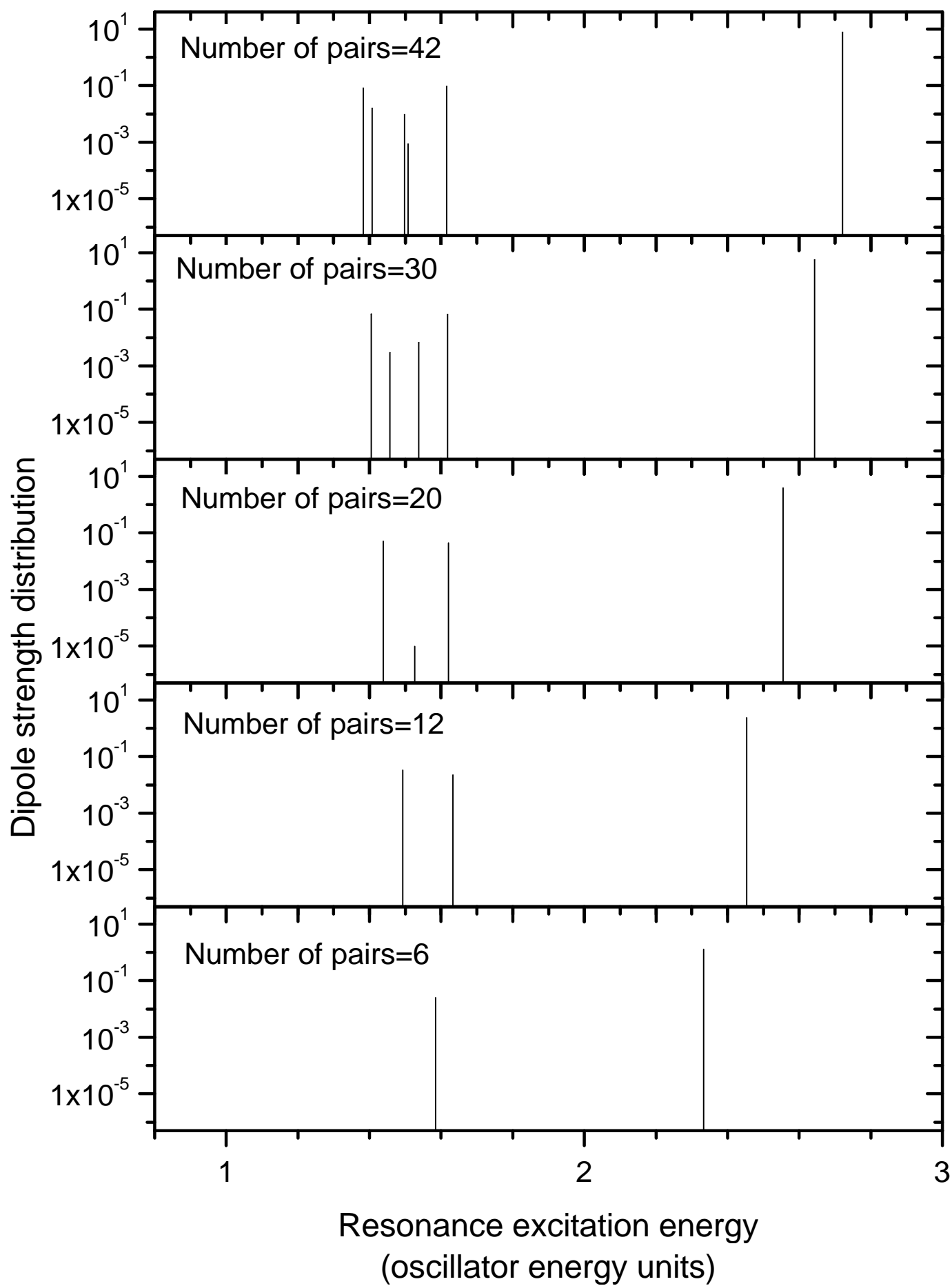
\end{document}